# Double Layer-Interlocked Crystals of Nitrogen-Rich Compounds under Zero-Pressure Conditions

Zhen Gong, [1,2,#], Baiqiang Liu [1,#], Xinrui Yang [1,2], Hongbo Jing [1,2], Ruiqi Xu [1,2] and Zhigang Wang [1,2,*]

[1] Key Laboratory of Material Simulation Methods & Software of Ministry of Education, College of Physics, Jilin University, Changchun, 130012, China

[2] Institute of Atomic and Molecular Physics Jilin University, Changchun, 130012, China

[#]The authors contributed equally to this work.

*Corresponding author. Email: wangzg@jlu.edu.cn (Z. W.)

**Abstract:**

Stabilizing nitrogen-rich compound crystals under conventional conditions is a key issue in the development and application of high–energy density materials (HEDMs). Herein, a two-dimensional double–layer interlocked $Li_4(N_5)_2$ nitrogen–rich compound crystals, in which the two $N_5$ rings are locked to by sharing four Li atoms, was found to maintain structural stability at zero pressure conditions. Dynamics studies reliably confirm crystal stability below 250 K. Furthermore, the stability of $Li_4(N_5)_2$ crystal mainly arises from the ionic interaction between Li atoms and $N_5$ rings, formed by the charge transfer from Li atoms to $N_5$ rings. This study highlights the feasibility of stabilizing nitrogen-rich compound crystals under conventional conditions, paving the way for atomic level advancements in HEDMs.

**Introduction**

High–energy density materials (HEDMs) are potential candidates for use in explosives, propellants and clean energy, and primarily include traditional nitro compounds, nuclear energy materials, hydrogen energy producing materials and nitrogen–rich compounds. As fourth–generation of HEDMs, nitrogen–rich compounds possess a high nitrogen content, with nitrogen atoms connected via single bonds or double bonds, enabling the release of substantial energy during of N–N bond transformations. Their decomposition product, $N_2$, is environmentally friendly, generating widespread interest. These compounds have been successfully synthesized and stabilized, under high pressure conditions (*1-6*). Recently, researchers have been focusing on achieving the stability of nitrogen-rich compounds under conventional pressure conditions for synthesizing new HEDMs.

How to achieve the stability of nitrogen-rich compounds under conventional pressure conditions is considered an important research direction for synthesizing new HEDMs. Currently, metal salt hydrates containing $N_5^-$, such as $(N_5)_6(H_3O)_3(NH_4)_4Cl$, have been successfully synthesized in acidic solution (*7*) and can remain stable in complex multi–component crystals, though at the cost of reduced energy density. To pursue high–energy density, numerous studies on nitrogen–rich compound crystals have been conducted, but they all need stability under high pressure. Therefore, stabilizing nitrogen–rich compound crystals under conventional conditions faces an important challenge. However, if we transform our perspective from crystals to molecules (*8, 9*), we find that some nitrogen–rich compound molecules are be stable under conventional conditions (*10-12*), showing a bottom–up potential for the synthesis of stable nitrogen–rich compound crystals under conventional conditions.

Notably, the crystals of nitrogen–rich compounds identified at high pressures are almost exclusively three–dimensional. Despite there being no evidence suggesting that this three–dimensional structure prevents the stabilization of nitrogen–rich compound crystals at conventional pressures, recent discoveries of exotic 2D crystals, such as graphene (*13-15*), transition metal dichalcogenides [16-18]. hexagonal boron nitride (h–NB) (*19, 20*), and layered metal oxides (*21, 22*), have been shown stability under conventional conditions. Based on this special crystal structure, a 2D double–layer crystal structure akin to ice (*23-28*), is proposed. Therefore, as the research into low–dimensional crystalline materials deepens has gradually emerged (*14-16, 29*). The discovery of 2D double–layer structures may offer a pathway for stabilizing nitrogen–rich compound crystals of HEDMs under conventional pressure.

Herein, we report a 2D double layer–interlocked $Li_4(N_5)_2$ nitrogen–rich compound crystal that is be stable under conventional and even zero–pressure conditions. Atomic level mechanism analysis clearly shows that the stability of this double–layer interlocking structure is owing to ionic interactions resulting from charge transfer between the interlayer Li atoms and $N_5$ rings. This discovery not only provides a promising approach for stabilizing nitrogen–rich compound crystals under conventional conditions, especially under normal pressure, but also open new avenues for the design and synthesis of stable HEDMs.

## Results and Discussion

To obtain a stable crystal structure of a nitrogen–rich compound under zero–pressure conditions, we first attempt to assemble the known stable $LiN_5$ molecules (*11*) in a one–dimensional direction (Fig. 1(a)). Using this method to generate initial guess structures, we performed numerous high–precision structural optimizations without pressure conditions along 2D directions during periodic structural relaxation (for details, see Method). Notably, theoretical calculations confirmed the lowest energy crystal structure $Li_4(N_5)_2$, which features a unique 2D double layer–interlocked configuration. As shown in Fig. 1(b) and 1c, the lattice constants of the primitive cell are a = 5.95 Å, b = 5.13 Å, c = 14.84 Å, α = 89.63°, β = 90.84° and γ = 88.83° (for structural parameters, see Supplementary Information (SI) TABLE S1).

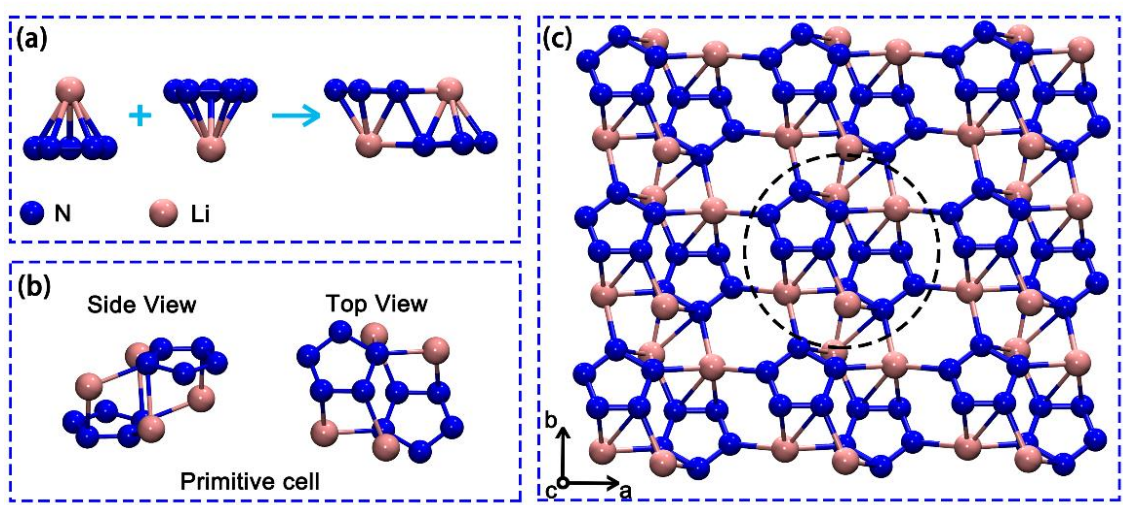

**Figure 1.** Geometric structures of the $Li_4(N_5)_2$ crystal. **(a)** One–dimensional assembly diagram of $LiN_5$. **(b)** Side and top views of the primitive cell of the crystal. **(c)** Supercell structure of the crystal, with the primitive cell marked by a dotted circle.

To verify the stability of 2D double layer–interlocked $Li_4(N_5)_2$ nitrogen–rich compound crystal, we performed phonon dispersion spectrum calculations and dynamics simulations of first–principles. As shown in Fig. 2a, the phonon dispersion spectrum indicates that all frequencies within the first Brillouin zone are positive, indicating that the mechanical stability of the crystal structure. Furthermore, we confirmed the stability of its structure through molecular dynamics simulation. Specifically, NVT ensemble simulations were run at each 50 K within the temperature range of 0–300 K, with each simulation lasting 10 ps. By performing sampling simulations at multiple temperature points, we discovered that the crystal structure maintained its basic 2D double layer–interlocked structure even when the temperature was raised to 250 K, indicating its thermal stability (see SI, Fig. S1). Combining these results, the phonon dispersion spectrum and dynamics simulations confirm that the two–site interlocking structure exhibits excellent stability.

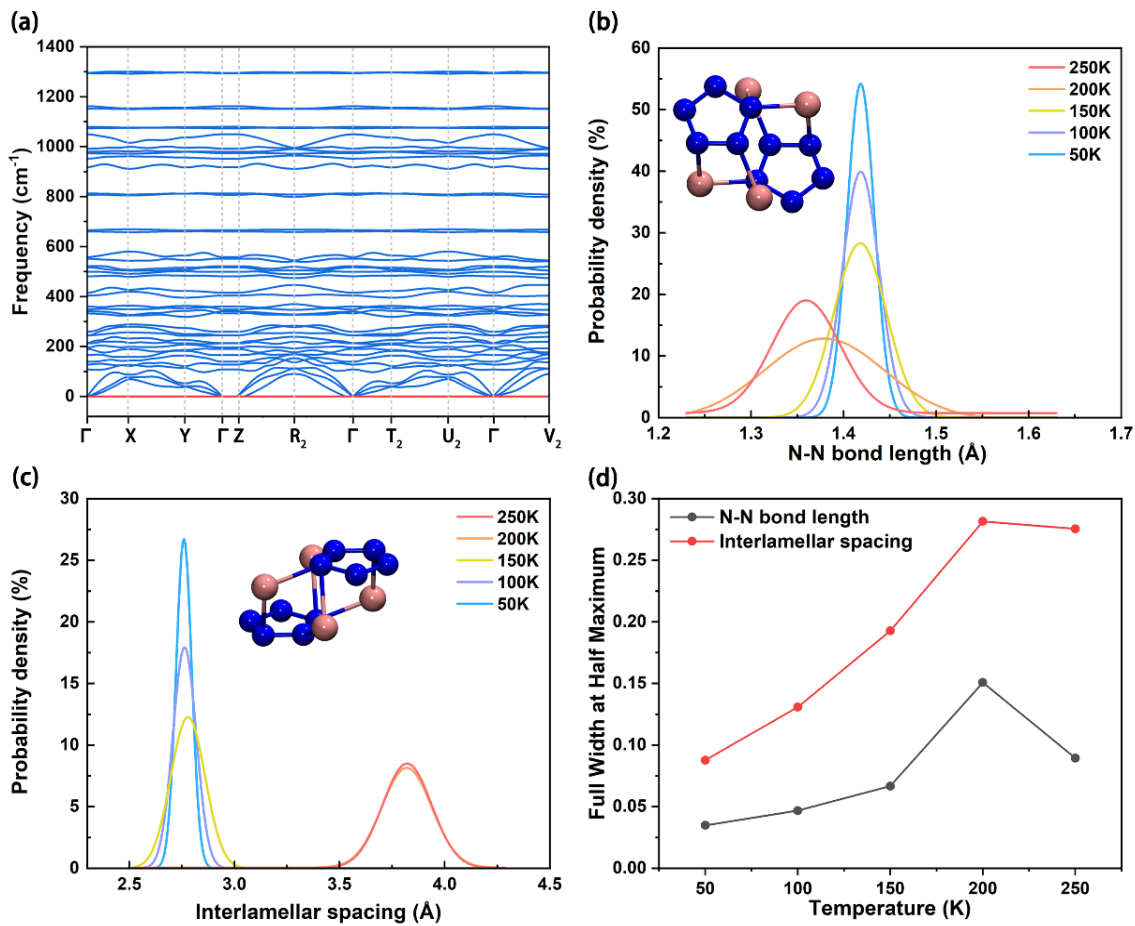

**Figure 2.** Phonon dispersion curve of the $Li_4(N_5)_2$ crystal and probability distribution of structural parameter changes from dynamic simulations at different temperatures. **(a)** Phonon dispersion curves of the crystal structure. **(b)** Distribution of N–N bond lengths within the $N_5$ ring of the crystal structure. **(c)** Distribution of interlayer spacing between $N_5$ rings in the crystal structure. **(d)** Full width at half maximum of N–N bond lengths and interlayer spacing between $N_5$ rings.

Based on the dynamic simulation results, we further statistically analyzed the N–N bond length and the interlayer spacing between $N_5$ rings in $Li_4(N_5)_2$ crystals at different temperatures. As shown in Fig. 2(b), the N–N bond lengths at different temperatures are predominantly distributed between 1.35 and 1.42 Å with a high probability. This indicates that the N–N bond length of the $N_5$ ring mainly varies within this range. Notably, the N–N bond of the $N_5$ ring in the crystal is considerably longer than a typical double bond length (1.20 Å) (*30*) but shorter than an average single bond (1.45 Å) (*31*), indicating that the $N_5$ ring in the crystal adopts a bonding mode between double and single bonds, consistent with previous theoretical predictions for nitrogen–rich compound crystals (*32*). Second, for the $Li_4(N_5)_2$ crystal structure, as the temperature increases, the interlayer spacing between $N_5$ rings is between 2.70 and 3.85 Å (Fig. (2c)), which aligns with the well–known stacking effect (*33*). Furthermore, an analysis of the structural energy changes over time during the dynamic simulation (see SI Fig. S2 for details), reveals that the energy change range of the system fluctuated around 0.2 eV, which further confirmed the thermal stability of the crystal. Finally, by statistically analyzing the changes in the full width at half maximum (FWHM) of the bond length and interlayer spacing. As shown in Fig. 2d (see also SI, Fig. S3 and S4), the full width at half maximum of the N–N bond length and interlayer spacing of the N$_5$ ring increases with temperature, suggesting great structural variation at elevated temperatures. The results indicate that the Li$_4$(N$_5$)$_2$ nitrogen–rich compound crystal maintains its 2D double layer–interlocked structure and stability below 250 K.

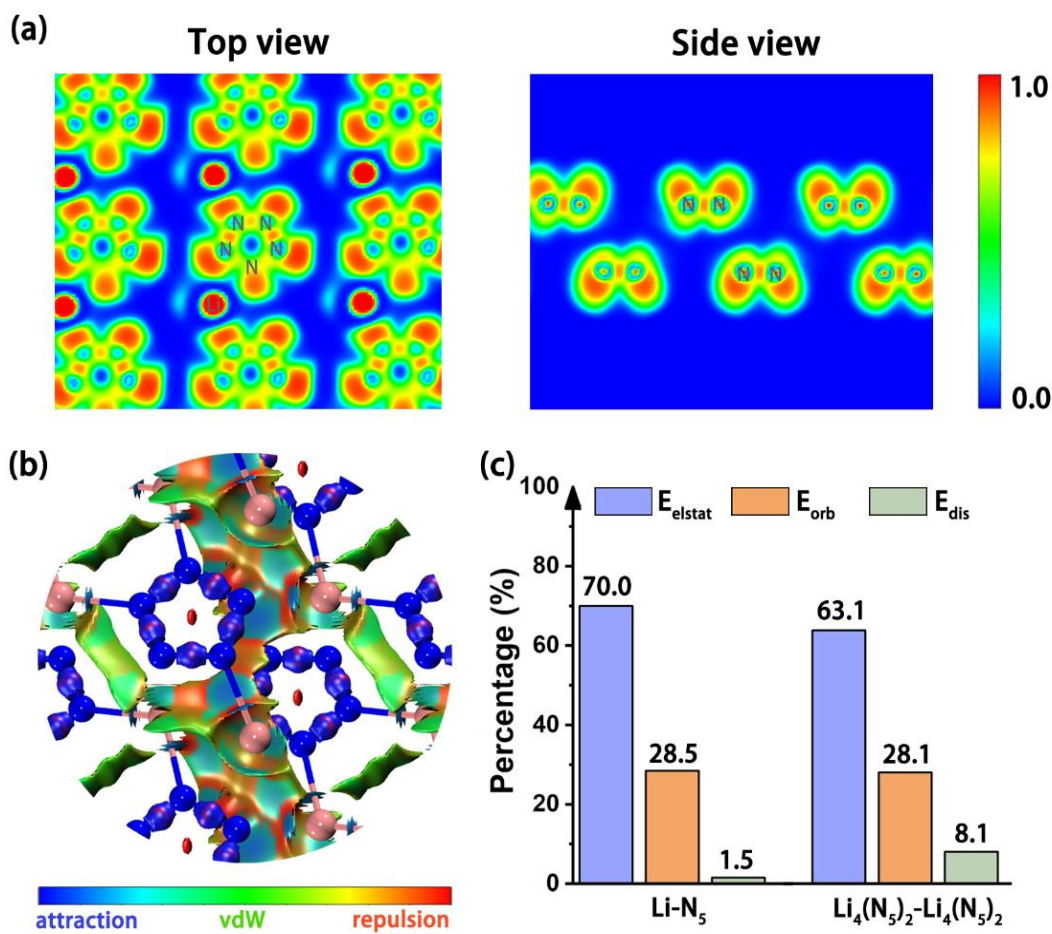

**Figure 3.** Interaction analyses within and between Li$_4$(N$_5$)$_2$ primitive cells. **(a)** Side view and top views of the electron localization function maps, where value 0.0 and 1.0 represent no localization (blue) and full localization (red), respectively. **(b)** Interaction region indicator, with isosurfaces in blue, green, and red representing attractive, van der Waals (vdW), and repulsive interactions, respectively. **(c)** Energy decomposition analysis (EDA) between Li atoms and N$_5$ rings within the primitive cell, as well as between Li$_4$(N$_5$)$_2$ in the primitive cell.

Building on the stable conformation of Li$_4$(N$_5$)$_2$, we conducted atom charge analysis and bond order analysis to explore the stability mechanism of this crystal (see TABLEs S2 and S3 in the SI for details). The Hirshfeld [34] and Mulliken [35] charge analyses reveal that electrons from the Li atom are transferred to the N$_5$ ring to form an anionic state. In addition, Mayer bond order [36] analysis shows no bonding between the Li atom and the N$_5$ rings, which further confirms that only charge transfer occurs between them. Therefore, the stability of the Li$_4$(N$_5$)$_2$ crystal is attributed to charge transfer between the Li atom and the N$_5$ rings

To understand how the crystal mechanism influences the stability of the unique 2D double layer–interlocked structure, we conducted an interaction analysis of the nitrogen-rich compound crystal Li$_4$(N$_5$)$_2$. First, the electron localization function (ELF) (*37*) analysis was employed to examine electron distribution within and between the primitive cells. As shown in Fig. 3(a), the blue color between the Li atom and the $N_5$ rings indicates that no covalent bond is formed between them. Combined with the atomic charge analysis, this suggests that ionic bonds form between the Li atom and the $N_5$ rings, with Li atoms acting as bridges connecting the two layers of $N_5$ rings. Thus, the primitive cell is interconnected by Li atoms.

To further confirm that the interior and exterior of the unit cells are connected by Li atoms, an interaction region indication (IRI) (*38*) analysis was performed. As shown in Fig. 3(b), within the primitive cell, the Li atom and the $N_5$ rings possess attraction, while van der Waals (vdW) interactions occur between the $N_5$ rings. In addition, between the primitive cells, the Li atom and adjacent $N_5$ rings possess attraction, with the $N_5$ rings exhibiting vdW interactions. These results further support the conclusion that the interaction between Li atom and $N_5$ rings in $Li_4(N_5)_2$ crystals is key to maintaining its stability. Additionally, energy decomposition analysis (EDA) (*39*) was performed to confirm the interaction properties between the Li atom and $N_5$ rings. As shown of Fig. 3c, the electrostatic interaction between the Li atom and the $N_5$ rings within primitive cell is dominant. Moreover, the electrostatic interaction between the primitive cells is also predominant. Therefore, this result indicates that the electrostatic interaction between the Li atom and the $N_5$ rings, both within and between the primitive cells, contributes to the formation of a 2D double layer–interlocked structure of the crystal. The consistent ELF, IRI and EDA analyses not only confirm the type of interaction between N–N and Li–$N_5$, but also strongly demonstrate the stability of the crystal structure.

**Summary**

In conclusion, we have been proposed a 2D double layer–interlocked nitrogen–rich compound crystal for the first time. Based on first–principles calculations, we determined that it can maintain both structural and dynamics stability under zero–pressure conditions. Further electronic structure analysis revealed that its stability mechanism comes from ionic interactions dominated by charge transfer between metallic lithium atoms and $N_5$ rings, facilitating the formation of a 2D double layer-interlocked structure. In summary, the discovery of this 2D double layer–interlocked nitrogen–rich compound crystal will pave the new way for obtaining HEDMs under conventional conditions.

**Method**

The $Li_4(N_5)_2$ crystal was optimized and analyzed using first-principle density functional theory (DFT) calculations. The exchange–correlation functional was treated using generalized gradient approximation (GGA) (*40*) within the Perdew–Burke–Ernzerhof (PBE) (*41*) functional. To optimize the primitive cell of the $Li_4(N_5)_2$ crystal, a cut off energy of 400 eV was set, and the Brillouin zone was represented by a 6 × 4 × 1 k mesh. The phonon calculations for crystal structure were performed using the supercell approach (*42*) as implemented in the phonopy code (*43*). All the above calculations used the CP2K (*44*) program, while atomic charge, Mayer bond order (*36*), sobEDA (*39*) and ELF (*37*) analyses were conducted using the Multiwfn 3.8 package (*45*).

**ACKNOWLEDGMENTS**

The authors wish to acknowledge Rui Liu, Rui Li and Chenxi Wan for discussion. This work was supported by the National Natural Science Foundation of China (Grant number 11974136). Z.W. also acknowledges the High-Performance Computing Center of Jilin University and the National Supercomputing Center in Shanghai.

# Supporting Information for

# Double Layer-Interlocked Crystals of Nitrogen-Rich Compounds under Zero-Pressure Conditions

Zhen Gong [1, 2, #], Baiqiang Liu [1, #], Xinrui Yang [1, 2], Hongbo Jing [1, 2], Ruiqi Xu [1, 2] and Zhigang Wang [1, 2, *]

[1] Key Laboratory of Material Simulation Methods & Software of Ministry of Education, College of Physics, Jilin University, Changchun, 130012, China

[2] Institute of Atomic and Molecular Physics Jilin University, Changchun, 130012, China

[#]The authors contributed equally to this work.

[*]*Corresponding author. Email: wangzg@jlu.edu.cn (Z. W.)*

**Tables**

**Table S1.** Calculated structural parameters of the $Li_4(N_5)_2$ crystal structure.

|  | Lattice Parameters (Å, °) | Atomic coordinates (fractional) Atoms | X | Y | Z |
|---|---|---|---|---|---|
| $Li_4(N_5)_2$ | a = 5.9514 | N | 2.86484 | 3.85428 | 8.24475 |
|  | b = 5.1316 | N | 2.43446 | 2.63292 | 8.75143 |
|  | c = 14.8355 | N | 1.08441 | 2.64212 | 8.78160 |
|  | α = 89.6270 | N | 1.66174 | 4.55682 | 7.90762 |
|  | β = 90.8406 | N | 6.37197 | 3.76268 | 8.27945 |
|  | γ = 88.8260 | N | 4.22884 | 6.47981 | 6.75361 |
|  |  | N | 3.02497 | 1.34851 | 6.41489 |
|  |  | N | 3.45386 | 2.57031 | 5.90841 |
|  |  | N | 5.25253 | 1.44325 | 6.38159 |
|  |  | N | 4.80394 | 2.56300 | 5.87882 |
|  |  | Li | 1.13944 | 1.24974 | 7.17692 |
|  |  | Li | 3.17376 | 7.93506 | 8.55661 |
|  |  | Li | 2.71663 | 4.41017 | 6.10233 |
|  |  | Li | 4.75108 | 3.95311 | 7.48395 |

**TABLE S2.** The calculated charges of Li$_4$(N$_5$)$_2$ crystal structure by the Hirshfeld population analysis and the Mulliken charges.

| Atom | Hirshfeld | Mulliken |
|---|---|---|
| **1(N)** | -0.14 | -0.25 |
| **2(N)** | -0.08 | -0.15 |
| **3(N)** | -0.05 | -0.14 |
| **4(N)** | -0.13 | -0.22 |
| **5(N)** | -0.07 | -0.13 |
| **6(N)** | -0.13 | -0.22 |
| **7(N)** | -0.14 | -0.25 |
| **8(N)** | -0.08 | -0.15 |
| **9(N)** | -0.07 | -0.13 |
| **10(N)** | -0.05 | -0.14 |
| **11(Li)** | 0.20 | 0.42 |
| **12(Li)** | 0.28 | 0.48 |
| **13(Li)** | 0.28 | 0.48 |
| **14(Li)** | 0.20 | 0.42 |

**TABLE S3.** The Mayer bond order of Li$_4$(N$_5$)$_2$ crystal structure.

| Mayer bond order | | | |
|---|---|---|---|
| 1(N)-2(N) | 1.18 | 4(N)-13(Li) | 0.19 |
| 1(N)-3(N) | 0.12 | 5(N)-7(N) | 0.07 |
| 1(N)-4(N) | 0.96 | 5(N)-14(Li) | 0.27 |
| 1(N)-5(N) | 0.05 | 6(N)-7(N) | 0.96 |
| 1(N)-6(N) | 0.09 | 6(N)-8(N) | 0.05 |
| 1(N)-7(N) | 0.07 | 6(N)-9(N) | 1.22 |
| 1(N)-9(N) | 0.07 | 6(N)-10(N) | 0.08 |
| 1(N)-12(Li) | 0.19 | 6(N)-12(Li) | 0.19 |
| 1(N)-13(Li) | 0.08 | 6(N)-13(Li) | 0.12 |
| 1(N)-14(Li) | 0.22 | 6(N)-14(Li) | 0.26 |
| 2(N)-3(N) | 1.43 | 7(N)-8(N) | 1.18 |
| 2(N)-4(N) | 1.41 | 7(N)-9(N) | 0.05 |
| 2(N)-5(N) | 0.18 | 7(N)-10(N) | 0.12 |
| 2(N)-12(Li) | 0.29 | 7(N)-11(Li) | 0.23 |
| 3(N)-4(N) | 0.08 | 7(N)-12(Li) | 0.08 |
| 3(N)-5(N) | 1.41 | 7(N)-13(Li) | 0.19 |
| 3(N)-11(Li) | 0.14 | 8(N)-9(N) | 0.18 |
| 4(N)-5(N) | 1.22 | 8(N)-10(N) | 1.43 |
| 4(N)-6(N) | 0.10 | 8(N)-13(Li) | 0.29 |
| 4(N)-7(N) | 0.09 | 9(N)-10(N) | 1.41 |
| 4(N)-11(Li) | 0.26 | 9(N)-11(Li) | 0.27 |
| 4(N)-12(Li) | 0.12 | 10(N)-14(Li) | 0.14 |

**Figures**

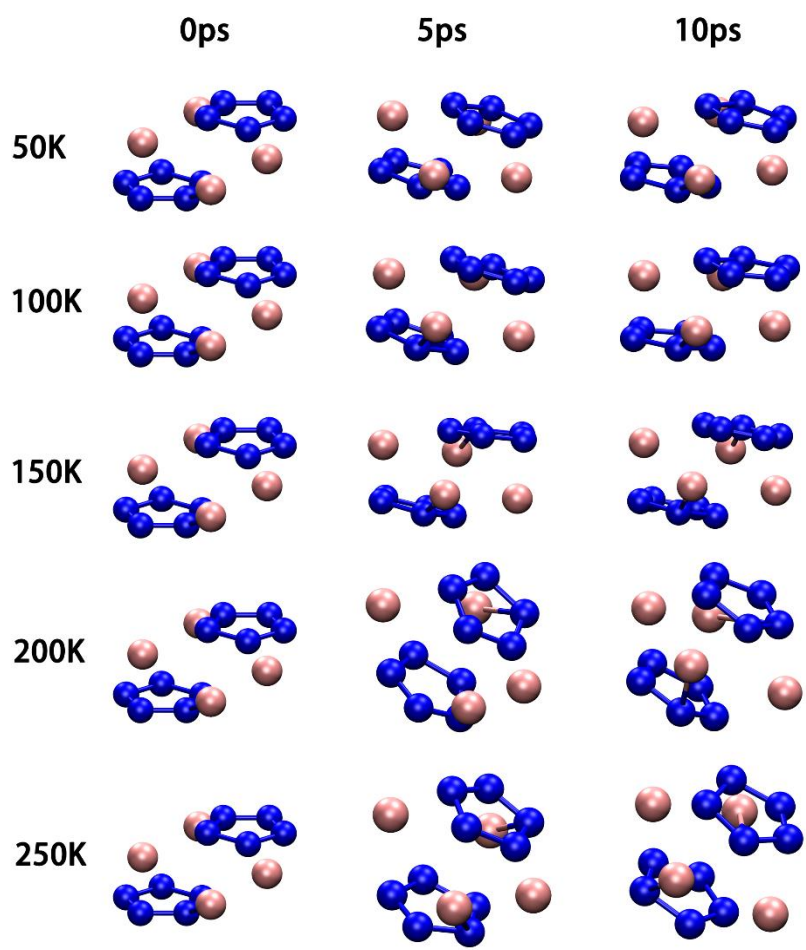

**FIG. S1.** The potential energy surface for chiral inversion of isolated thalidomide molecule. The solid lines represent channel a and the dashed lines represent channel.

To verify the structural stability of $Li_4(N_5)_2$ crystal structure, ab initio molecular dynamics (MD) simulations were conducted. The simulations were performed at a temperature of 50 K to 250 K, with a simulation time duration of 10 ps and a time step of 1 fs.

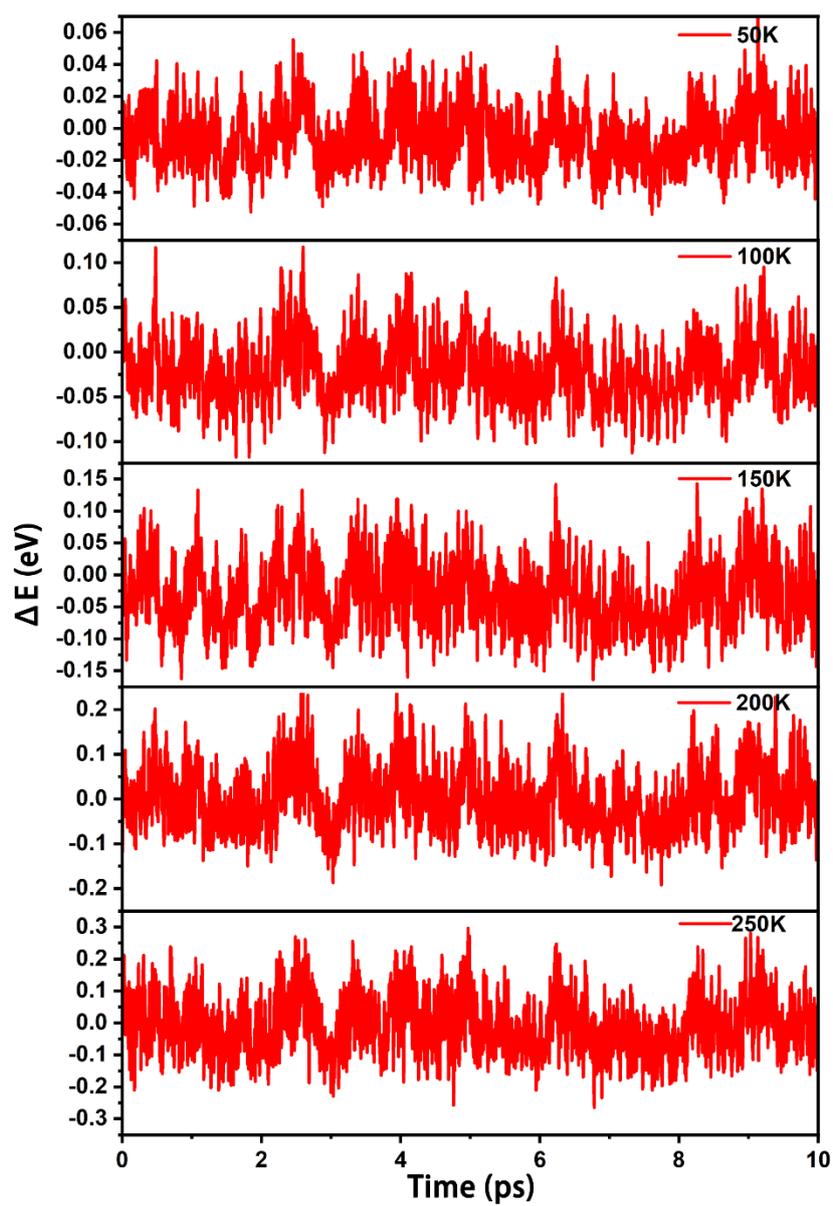

**FIG. S2.** Variations of total energy as a function of dynamic simulation time from 50 K to 250 K.

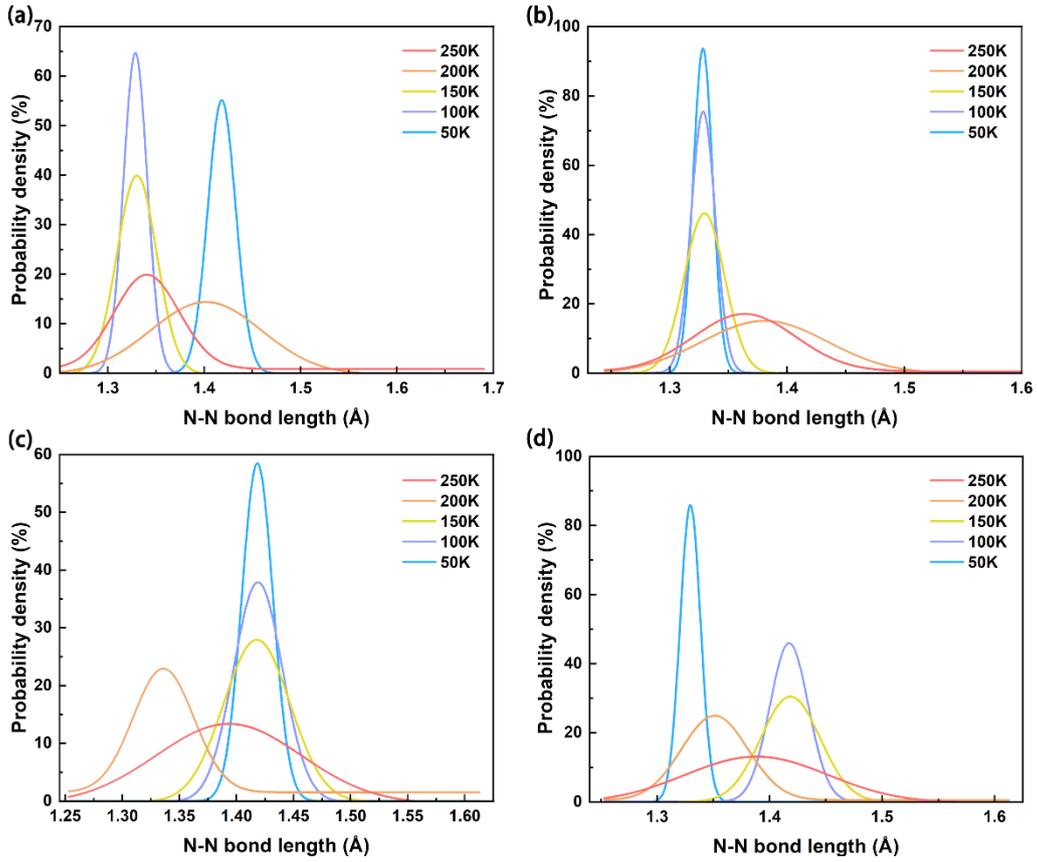

**FIG. S3.** Distribution of N-N bond lengths of the $N_5$ ring in the crystal structure.

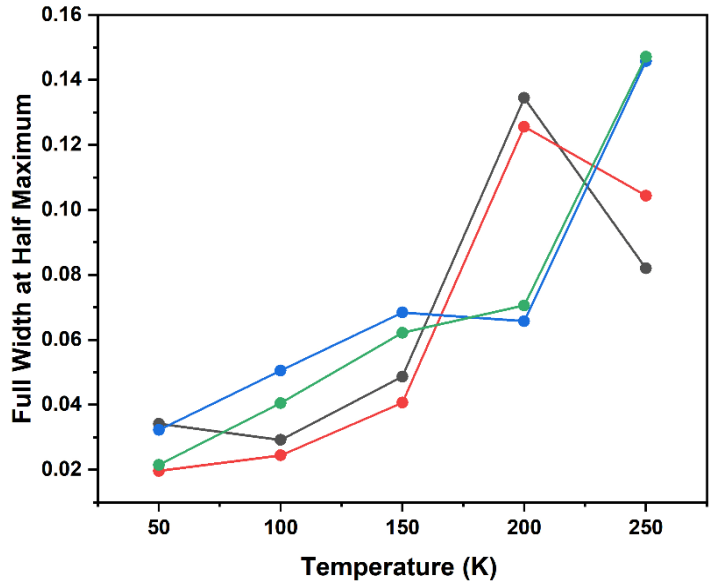

**FIG. S4.** The Full Width at Half Maximum (FWHM) of N-N bond lengths and interlayer spacing between $N_5$ rings.
TABLE S2. The calculated charges of $Li_4(N_5)_2$ crystal structure by the Hirshfeld population analysis and the Mulliken charges.

13